# Exception Agent Detection System for IP Spoofing Over Online Environments

Al-Sammarraie Hosam
Center for IT and Multimedia, Universiti
Sains Malaysia
Penang, Malaysia

Adli Mustafa
School of Mathematical sciences, Universiti
Sains Malaysia
Penang, Malaysia

Shakeel Ahmad
School of Mathematical sciences, Universiti
Sains Malaysia, Institute of Computing and
Information Technology, Gomal University,
Pakistan
Penang, Malaysia
.

Merza Abbas
Center for IT and Multimedia, Universiti
Sains Malaysia
Penang, Malaysia

*Abstract*—Over the recent years, IP and email spoofing gained much importance for security concerns due to the current changes in manipulating the system performance in different online environments. Intrusion Detection System (IDS) has been used to secure these environments for sharing their data over network and host based IDS approaches. However, the rapid growth of intrusion events over Internet and local area network become responsible for the distribution of different threats and vulnerabilities in the computing systems. The current signature detection approach used by IDS, detects unclear actions based on analyzing and describing the action patterns such as time, text, password etc and has been faced difficulties in updating information, detect unknown novel attacks, maintenance of an IDS which is necessarily connected with analyzing and patching of security holes, and the lack of information on user privileges and attack signature structure. Thus, this paper proposes an EADS (Exception agent detection system) for securing the header information carried by IP over online environments. The study mainly concerns with the deployment of new technique for detecting and eliminating the unknown threats attacks during the data sharing over online environments.

## I. INTRODUCTION

The rapid growth of intrusion events over local area network and Internet have been distributed among the organizations and other environments, which pushed most of these environments to implement security techniques against corresponding threats. Internet Protocol (IP) provides sustainable services for information delivery across Internet. The packet will present these information depends on TCP/IP layers. IP datagram contain a header for caring the source details of network to be forwarded to the same IP datagram destination. Details carried by IP header are a) time to live b) source and destination addresses c) types of service and others relevant information. The importance of header to send and receive information over LAN is usually grabbed by attackers. Moreover, attackers may also use some other techniques to grab the header information carried via IP over LAN online environments [5].

Hence environments used to integrate powerful techniques for detecting and preventing IP changes, such as intrusion detection system (IDS) may deploy to secure and monitor IP behavior over LAN.

Intrusion Detection Systems are tools to assist in managing threats and vulnerabilities in this changing environment. Threats are people or groups who have the potential to compromise other computer system [4]. These may be an inquisitive teenager, a discontented worker / employee, or spy from an opponent company or any foreign government. Attacks on network computer system could be devastating, affect networks, and corporate establishments. It's requiring to provide and curb these attacks and Intrusion Detection System helps to identify the intrusions. Without an NIDS, to monitor any network activity, possibly resulting in irreparable damage to an organization's network. Intrusion attacks are "those attacks in which an attacker enters your network to read, damage, and/or steal your data" [1]. These attacks can be divided into two subcategories: pre intrusion activities and intrusions.

The current IDS used two intrusion detection approaches firstly; anomaly detection approach, that used to manipulate the relation between profile and the current behavior of the TCP/IP, also determine the difference between profiles and detect possible attack attempts. Secondly; signature detection approach, used to detect ambiguous and unclear actions by analyzing and describing the action patterns such as (time, text, password etc) [11].

Figure 1 shows the proposed EADS workflow over online environments such as (ENA, ENB, ENC, and END). Data sharing over these environments presented the IDS technique for securing IP datagram during the transfer between environments to another. However, many protocols and architectures for LAN and Host based IDS were designed without taking care of the possibility of other threats attack. Moreover, the existing defense mechanisms against such attacks in host network are not effective to analyze and laminate the unknown attacks, which back to the differences in their characteristics. Furthermore, most of these environments require secure systems to detect and eliminate the external and internal attacks from other attackers over LAN and Host based IDS.

Environments are constantly evolving and changing with the emergence of new technologies and the Internet which may introduce new threats and unknown attacks. Hence, intrusion detection system has been used to secure and support these environments for sharing their data over LAN and Host based IDS. Furthermore, IP spoofing over online environments presents different patterns and follow exceptional behaviors based on attacker's techniques. This paper mainly concerns with the deployment of new technique for detecting and eliminating the unknown threats attacks during the data sharing.

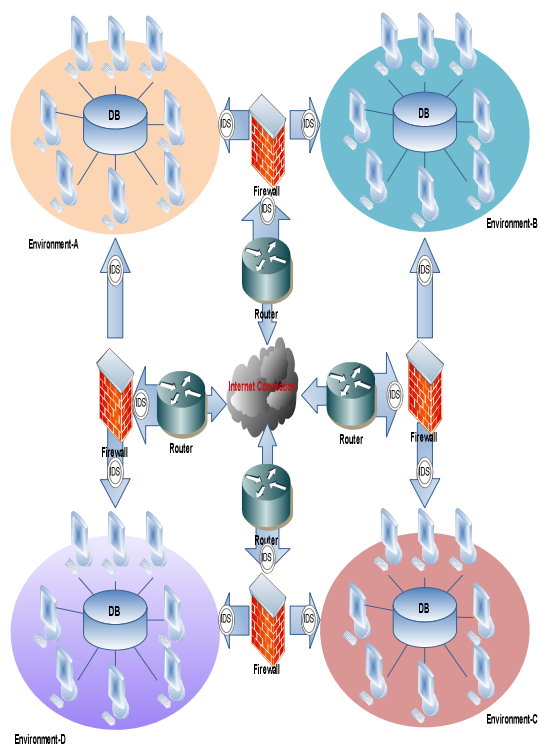

Figure 1. Online environment workflow

Additionally, different disadvantages have been detected over using the current signature detection approach such as (Difficulties in updating information, unable to detect unknown novel attacks, maintenance of an IDS is necessarily connected with analyzing and patching of security holes, the attack knowledge is operating environment dependent, and the lack of information on user privileges and attack signature structure). Moreover, there is a lack of detecting and describing the new IP spoofing patterns based on signature detection approach, such as a) random spoofing b) spoof a set of addresses consistently c) spoof a small address based on the attackers moves from set to another and etc. Hence, we proposed an improvement in the current detection techniques for IP spoofing over online environments based on signature detection approach.

Moreover, this paper proposes an enhancement in the current detection approach (signature approach) in terms of exception agent (virtual agent) system to examine and analyze the network traffic (packets) over multiple online environments. In addition, this paper will follow the classification of intrusion detection system which modified to present the study requirements.

Agent system is used to monitor and organize the networks performance among different threats over LAN. An agent is used to present the threats behavior for processing these behaviors over LAN and Host. Moreover, an agent is a computational system in which agents with different capabilities and resources perform their task by coordinating and cooperating with each other in order to achieve a set of goals through interaction, coordination, cooperation and collective intelligence [9].

Rest of the paper is organized as follows. Related work is presented in section 2. Section 3 presents IDS methodology. EADS is presented in section 4. Experimental results based on HIDS presented in section 5 followed by conclusion.

## II. RELATED WORK

Recently, different studies have been presented to describe the architecture and the implementation of techniques for detecting and manipulating the spoofing activities over LAN. However, researchers such as [3] explained the Probabilistic Agent-Based Intrusion Detection (PAID) system. These systems provide cooperative agent architecture, which can perform specific intrusion detection tasks (e.g., identify IP-spoofing attacks). PAID allow to other agents to share the probability distribution of an event occurrence.

A study presented a framework to investigate the prospective adaptive and cooperative defense mechanisms against the Internet attacks. The suggested approach is based on the multi agent modeling and simulation. This framework represents the attack as interacting teams of intelligent agents that act under some adaptation criterion. They adjust their configuration and behavior in compliance with the network conditions and attack (defense) severity [6].

However, a study reported the design and evaluation of the Clouseau system, with the route-based filtering (RBF). This design was an effective and provide practical defense against IP spoofing. Since RFB process critically customize on the accuracy of the IP layer information that used for spoofed packet detection. The inference process as described by them is "resilient to subversion by an attacker who is familiar with Clouseau" [8].

Another study proposed an ANTID scheme for detecting and filtering DDoS attacks which uses spoofed packets to circumvent the conventional intrusion detection schemes. This ANTID intends to complement the conventional schemes by embedding in each IP packet a unique path fingerprint that represents the route an IP packet has traversed; ANTID is able to distinguish IP packets that traverse different Internet paths [7].

A study presented a model and architecture for enhancing the current signature detection approach based on intrusion detection engine with different threat capability. They modeled and enhanced the efficacy of the threat-aware signature based intrusion detection approach for obtaining network specific useful alarms. Furthermore, the study presented its experiments based on various threat scenarios and the obtained results shown that external threats formed 95% of the alarms by using the proposed model [10].

## III. METHODOLOGY

An intrusion detection systems methodology (IDS) is concerned with the detection of hostile actions [2]. Moreover, this selected methodology will present two main techniques i.e. the first technique of anomaly detection in general investigates issues associated with contradiction/deviations from normal routine system/user behavior whereas the $2^{nd}$ technique employs signature detection approach use to distinguish between attack or anomaly signatures and known ID signatures. Figure 2 shows the classification interaction detection system.

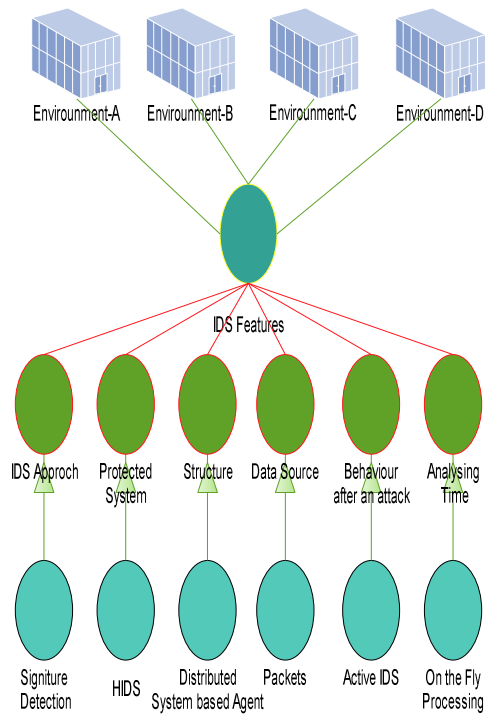

Figure 2. Classification of intrusion detection system modified version

There are different IDS tools for exploiting IDS information such as a) host based IDS (HIDS) which exploit host details from single host b) network based IDS (NIDS) which exploit IDS information from multiple signals of a local network. This paper presents EDS (Exception Detection System) technique to detect IP spoofing in HIDS. This study has been employed an exception agent detection system in HIDS environments.

## IV. EADS TECHNIQUE

A typical description of the process involved in exception agent detection technique is:

{

ENA ⟶ Environment A

ENB ⟶ Environment B}

*// selected environments to analyze and detect unknown IP spoofing//*

{

ARPS ⟶ IP Sent

ARPR ⟶ IP Received

ARPC ⟶ Examine the received IP with the Mack address data}

*// this part will present an examinatation of the ARP information (compare IP with existing Mack addresses) for data transfer from ENA to ENB //*

{

VARP ⟶ Virtual IP & Mack (Extract the ARP (IP) source from $EN_A$ and the received ARP (IP) from $EN_B$)

VGS ⟶ Virtual Agent (Send the same source ARP (IP) request from $EN_A$ to the unknown ARP (IP) received from $EN_B$)

VGR ⟶ Virtual Agent (Receive a request about ARP (IP) from $EN_B$)

VGC ⟶ Virtual Agent Compares and examines the extracted ARP (IP) source from ENA (VGS) with the ARP (IP) of VGR from ENB

}

*// finally, we will create a virtual ARP packet based source IP of ENA, to analyze and examine the unknown ARP received (ARPR) from ENB //*

*//*1, *2, and *3 presents the exceptional process for detecting the unknown threats over host networks//*

1 ENA send request to ENB **\*1**

2 Send ARPS from ENA to ENB

3 Receive ARPR request from ENB

4 Match ARPS from ENA with ARPR from ENB

5 ARPC = (ENA/ARPS) * (ENB/ARPR)

6 If ARPC > 0 **\*2**

7 Then save ARPC and transfer data from ENA to ANB

8 Else Create VARP

9 Extract ENA/ARPS & ENB/ARPR

10 Create VGS and VGR

11 Resend request VGS to ENB

12 (ENA/VGS) * (VGS/ VARP) = (ENA/VARP)

13 Receive VGS request from ENB

14 (ENB/VGR) * (VGR/VARP) = (ENB/VARP)

15 Compare based VGC = (ENB/VARP) * (ENA/ARPS)

16 If VGC > 0 **\*3**

17 Then save VGC in ARP

18 Else, IP spoofing Alarm

19 Stop transfer (Data sharing)

20 Eliminate VARP, VGS, VGR, and VGC

21 End

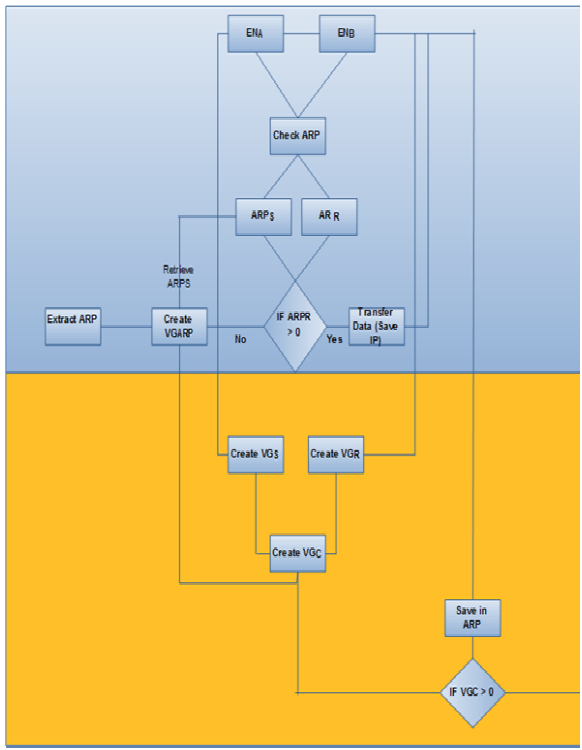

Figure 3. EADS activity diagram

## V. EXPECTED EXPERIMENTAL RESULTS

Figure 4 shown below is a simple process of the proposed technique based on HIDS.

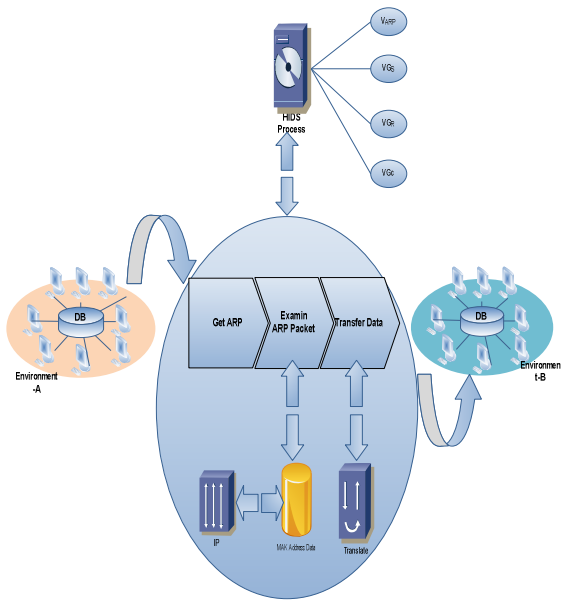

Figure 4. The EADS implementation architecture

This technique is used to examine and analyze the attempts threats over online environments. Additionally, the proposed EADS will deploy a virtual agent and virtual ARP based intrusion detection system during the attack from unknown IP, which lacks in the existing signature detection approach used by conventional IDS. These virtual agents will detect and analyze the unknown threats over host network by catching the unknown IP information and matching it with the existing IP source information. Then the proposed system will identify the unknown IP threats by inserting the IP information in the virtual ARP (during the attack) and save it later to the main ARP (after the attack).

Virtual ARP has been implemented to save the incoming IP address from one host environment to another in the host networks. Furthermore, the virtual ARP used to resolve and analyze different network layer protocol addresses to map hardware addresses, which it's primarily used to translate IP addresses to Ethernet MAC addresses.

This process will help to recognize and identify the unknown threats during the matching process of the incoming IP addresses over host networks with the existing IP information in the main ARP. Moreover the proposed model is expected to:

• Analyse and monitor user and system activities more efficiently.

• Provide much better system auditing and configuration vulnerabilities.

• Facilitates with better integrity of data and system files.

• Recognize pattern reflecting known attacks more smartly and provide efficient statistical analysis for ambiguous/abnormal activities.

• Efficiently monitor the data trail and tracing activities from start to exit point.

## VI. CONCLUSION

Nowadays, the rapid growth of designing and developing new techniques to secure data transferring over online environments have been deployed against certain network-oriented attacks like IP spoofing, packet storms, etc. that can be detected via IP datagram examination. This paper presents EADS, which deploy virtual agent based intrusion detection system during the attack from unknown IP, which lacks in existing IDS. The paper also presents process flow of the proposed model. The expected results presented in the paper shows the credibility of the proposed model.

ACKNOWLEDGMENT

This work is funded in part by USM Fellowship and USM RU grant.

AUTHORS PROFILE

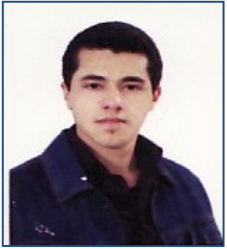

**Mr.Hosam Alsammarie** received his Bachelor Degree in Computer Engineering from Iraq (2006) and his Master in Infromation Technology from University Utara Malaysia (UUM). Curentlly, he is deoing his PhD by rsearch in University Saince Malaysia (USM) under followship schema. His research interests in Network Security, Ontology classification, Multi Agent System, Mobile programming based knowledge,and DBMS. He has presented papers in International conferences.

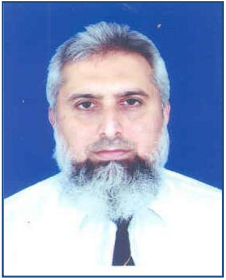

**Dr. Shakeel Ahmad** received his B.Sc. with distinction from Gomal University, Pakistan (1986) and M.Sc. (Computer Science) from Qauid-e-Azam University, Pakistan (1990). He served for 10 years as a lecturer in Institute of Computing and Information Technology (ICIT), Gomal University Pakistan.

Now he is serving as an Assistant Professor in ICIT, Gomal University Pakistan since 2001. He is among a senior faculty member of ICIT. Mr. Shakeel Ahmad received his PhD degree (2007) in Performance Analysis of Finite Capacity Queue under Complex Buffer Management Scheme.

Mr. Shakeel's research has mainly focused on developing cost effective analytical models for measuring the performance of complex queueing networks with finite capacities. His research interest includes Performance modelling, Optimization of congestion control techniques, Software Engineering, Software Refactoring, Network security, Routing Protocols and Electronic learning. He has produced many publications in Journal of international repute and also presented papers in International conferences.

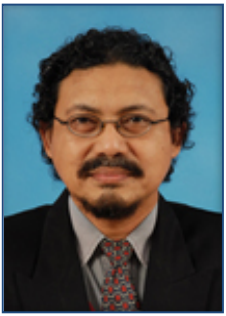

**Assoc. Prof. Dr. Merza Abbas** is working as a head researchers in center of IT and Multimedia, USM, Malaysia. He served for 30 years as a lecturer in University Saince Malaysia and other international insititutes. He has produced many publications in Journal of international repute and also presented papers in International conferences.